\documentclass[prl,aps,twocolumn,superscriptaddress,longbibliography]{revtex4-2}
\DeclareUnicodeCharacter{2212}{\textminus}

\usepackage{lineno}
\usepackage{hyperref}
\usepackage{amsmath}
\DeclareMathOperator{\sign}{sgn}
\usepackage{url}
\usepackage[caption=false,font={normalsize}]{subfig}
\usepackage{graphicx}
\usepackage{verbatim}
\usepackage{multirow}
\usepackage{amsmath, bm}
\newcommand{\beq}{\begin{eqnarray}}
\newcommand{\eeq}{\end{eqnarray}}
\usepackage{cleveref}
\usepackage{mathrsfs}
\usepackage{float}
\usepackage[usenames, dvipsnames]{color}
\usepackage{mathtools}
\usepackage{slashed}
\usepackage{physics}	
\usepackage{graphicx}   
\usepackage{epstopdf}
\usepackage{hyperref}   
\hypersetup{
pdfnewwindow=true, colorlinks=true,
linkcolor=blue, anchorcolor=blue,
citecolor=blue, filecolor=blue,
menucolor=blue, urlcolor=blue} 

\usepackage[export]{adjustbox}

\usepackage{bbold}
\usepackage{wasysym}
\usepackage{amssymb}
\definecolor{darkgray}{RGB}{10,30,30}
\definecolor{blue2}{rgb}{0.2, 0.2, 0.6}
\definecolor{blue3}{rgb}{0.16, 0.32, 0.75}
\definecolor{darkred}{rgb}{0.8,0,0}
\definecolor{royalblue}{rgb}{0.0, 0.14, 0.4}
\definecolor{magenta}{cmyk}{0,.9,0,0.2}
\definecolor{amethyst}{rgb}{0.6, 0.4, 0.8}
\definecolor{cadmiumgreen}{rgb}{0.0, 0.42, 0.24}
\definecolor{deepcarmine}{rgb}{0.66, 0.13, 0.24}
\definecolor{forestgreen}{rgb}{0.13, 0.55, 0.13}
\newcommand{\ncmd}{\newcommand}
\ncmd{\sect}[1]{\emph{\textbf{{#1}}}~---~}

\ncmd{\para}[1]{\paragraph*{{\color{black}{\bf #1:}}} }

\ncmd{\note}[1]{{\color{gray}{[\ding{168} #1}]}}

\ncmd{\MMnote}[1]{{\color{blue}{[\ding{168} {\bf #1}}]}}
\ncmd{\YWnote}[1]{{\color{purple}{[\ding{168} {\bf #1}}]}}

\ncmd{\sur}[1]{{\color{forestgreen}{ #1}}}
\ncmd{\qs}[1]{{\color{magenta}{ #1}}}
\ncmd{\qsnote}[1]{{\color{magenta}{ #1}}}

\newcommand{\beginsupplement}{
        \setcounter{table}{0}
        \renewcommand{\thetable}{S\arabic{table}}
        \setcounter{figure}{0}
        \renewcommand{\thefigure}{S\arabic{figure}}
        \setcounter{equation}{0}
        \renewcommand{\theequation}{S\arabic{equation}}
        \setcounter{section}{0}
        \renewcommand{\thesection}{\Alph{section}}
        \setcounter{subsection}{0}
        \renewcommand{\thesubsection}{\arabic{subsection}}
}

\ncmd{\yw}[1]{{\color{blue}{#1}}}
\begin{document}
\title{Suppression of shot noise at a Kondo destruction 
quantum critical point}
\author{Yiming Wang}
\affiliation{Department of Physics \& Astronomy,  Extreme Quantum Materials Alliance, Smalley-Curl Institute, Rice University, Houston, Texas 77005, USA}
\author{Shouvik Sur}
\affiliation{Department of Physics \& Astronomy,  Extreme Quantum Materials Alliance, Smalley-Curl Institute, Rice University, Houston, Texas 77005, USA}
\author{Fang Xie}
\affiliation{Department of Physics \& Astronomy,  Extreme Quantum Materials Alliance, Smalley-Curl Institute, Rice University, Houston, Texas 77005, USA}
\author{Haoyu Hu}
\affiliation{Department of Physics \& Astronomy,  Extreme Quantum Materials Alliance, Smalley-Curl Institute, Rice University, Houston, Texas 77005, USA}
\affiliation{Department of Physics, Princeton University, Princeton, NJ 08544, USA
}
\author{Silke Paschen}
\affiliation{Institute of Solid State Physics, Vienna University of Technology, Wiedner Hauptstr. 8-10, 1040, Vienna, Austria}
\affiliation{Department of Physics \& Astronomy,  Extreme Quantum Materials Alliance, Smalley-Curl Institute, Rice University, Houston, Texas 77005, USA}
\author{Douglas Natelson}
\affiliation{Department of Physics \& Astronomy,  Extreme Quantum Materials Alliance, Smalley-Curl Institute,
Rice University, Houston, Texas 77005, USA}
\author{Qimiao Si}
\affiliation{Department of Physics \& Astronomy,  Extreme Quantum Materials Alliance, Smalley-Curl Institute,
Rice University, Houston, Texas 77005, USA}

\begin{abstract}
Strange metal behavior has been observed in an expanding list of quantum materials, with heavy fermion metals serving as a prototype setting. Among the intriguing questions is the nature of charge carriers; there is an increasing recognition that the quasiparticles are lost, as captured by Kondo destruction quantum criticality.
Among the recent experimental advances is the measurement of shot noise
in a heavy-fermion strange metal.
We are thus motivated to study current fluctuations by advancing a minimal 
Bose-Fermi Kondo lattice model,
which admits a well-defined large-$N$ limit. Showing that the model in equilibrium captures
the essential physics of Kondo destruction, we proceed to
derive quantum kinetic equations and compute shot noise to the leading nontrivial order in
$1/N$. Our results reveal a strong suppression of the shot noise at the Kondo destruction quantum critical point, thereby providing the 
understanding of the striking experiment. Broader implications of our results are discussed.
\end{abstract}
\maketitle

For interacting many-electron systems
in dimensions higher than one, Fermi liquid theory prevails 
when the repulsive and short-ranged electron-electron 
interactions are treated perturbatively~\cite{Nozieres18}. 
In the regime of strong correlations, corresponding to the interactions reaching or exceeding the underlying electron bandwidth, the behavior becomes exceedingly rich~\cite{Kei17.1,Pas21.1}. 
Indeed, in a wide range of strongly
correlated electron systems, strange metals develop
~\cite{Phillips-science22,Hu-Natphys2024}, 
often in the vicinity of a quantum critical point 
(QCP)~\cite{Pas21.1,park_hidden_2006,nguyen_superconductivity_2021,legros_universal_2019,
greene_strange_2020,hayes_scaling_2016,zhang_high-temperature_2024}.
In addition to the linear-in-temperature resistivity, an emerging overall profile 
of strange metals includes such features as a loss of quasiparticles at the QCP, 
a jump of the Fermi surface across the QCP and a dynamical 
($\hbar \omega /k_{\rm B} T$) scaling~\cite{Hu-Natphys2024}.
All these salient
 features have been well established in heavy fermion strange metals~\cite{paschen2004,Friedemann.10,shishido2005,Schroder,Aro95.1,Prochaska2020,si2010,KirchnerRMP}, {and are the central properties of } 
Kondo destruction quantum criticality~\cite{Hu-Natphys2024,Si-Nature,Colemanetal,senthil2004a}. 
{These features reflect the physics of proximity to
electron localization-delocalization~\cite{Pas21.1,KirchnerRMP},
and}
have also been implicated to various degrees in other materials classes of strange metals. 
For example, a jump of the Fermi surface has been indicated in the 
 high-$T_c$ cuprates~\cite{Bad16.1}, 
 Fe-based superconductors~\cite{Hua22.1} and organic charge transfer salts~\cite{Oik15.1}.

Recently, a wave of novel experiments have provided new 
opportunities to deepen our understanding of strange metallicity~\cite{Savitsky2025}.
Among these are the shot noise measurements 
in the heavy fermion strange metal YbRh$_2$Si$_2$~\cite{Natelson2022}. The experiment revealed a striking suppression of the Fano factor -- a dimensionless quantity characterizing current fluctuations -- compared to 
{the expectations of both} conventional diffusive metals~\cite{Nagaev1992,Nagaev1995,Rudin1995,henny1999} 
and strongly correlated Fermi liquids~\cite{wang2022}. 
This observation suggests that strange metallicity manifests
beyond linear-response transport, extending into {(nonequilibrium)} current fluctuations.

The experiment in YbRh$_2$Si$_2$
has motivated a range of efforts to understand the shot noise in non-Fermi liquid systems
based on 
{both} general considerations and 
microscopic model calculations~\cite{Natelson2022,wang2022,Nikolaenko2023,wang2024,tc2024,wu2024}.
What has so far been lacking is 
any theoretical study of shot noise that captures the aforementioned overall profile of strange metallicity. 
YbRh$_2$Si$_2$ is one of the canonical heavy fermion metals 
that manifests the salient features in the overall profile. 
For example, the Fermi surface jump across 
its QCP has been established based on extensive measurements of an isothermal evolution of the Hall effect~\cite{paschen2004,Friedemann.10}
 (for recent reviews, see Refs.\,\onlinecite{Pas21.1,KirchnerRMP}).
The understanding of such features from the 
Kondo destruction QCP has in particular come 
from analyzing the competition between the Kondo and RKKY interactions 
via an extended dynamical mean field theory (EDMFT)~\cite{Hu-Natphys2024,lcp-prb}. 
Within EDMFT,  the local correlation functions of the Kondo lattice are 
determined in terms of a self-consistent Bose-Fermi Kondo model,
which describes couplings of the local moment
to both fermionic and bosonic baths.  
The EDMFT approach has successfully accounted for 
key experimental observations in quantum critical heavy fermion metals, 
and predicted features such as the Fermi surface jump~\cite{Hu-Natphys2024}.
However, formulating the EDMFT approach in nonequilibrium
settings is a major methodological challenge.
Hence, it is necessary to develop an alternative approach that incorporates 
{both} the essential physics of Kondo destruction 
{in equilibrium} and
the ability to treat nonequilibrium transport.

In this paper, we do so by advancing a minimal 
model -- the Bose-Fermi Kondo lattice model (BFKLM).
With the spin symmetry extended from $SU(2)$ to $SU(N)$, the 
 model 
admits a 
{well-defined} large-$N$ limit that enables analytical access to the quantum critical behavior 
associated with the Kondo destruction.
We employ Keldysh formalism to 
{construct} the quantum kinetic equations and compute the shot noise to the leading nontrivial order in the $1/N$ expansion. Our results reveal a pronounced suppression of the shot noise.
In this way, they provide the understanding of the experimental observations in YbRh$_2$Si$_2$ 
{and, by extension, uncover much-needed new insights into strange metallicity.}

\begin{figure}[!t]
\centering
\includegraphics[width=\columnwidth]{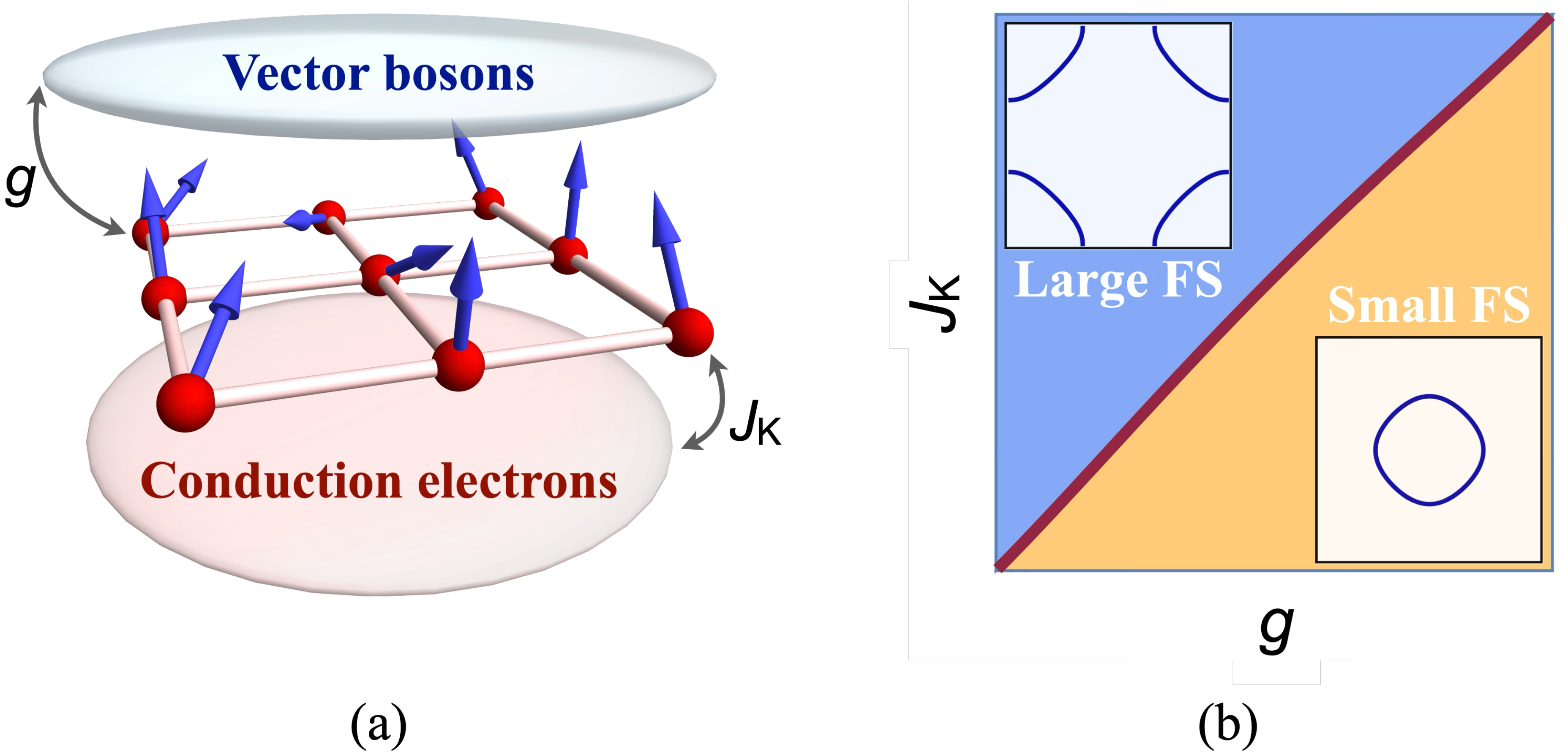}
\caption{
(a) A lattice illustration of the model, BFKLM, defined in Eq.\,(\ref{ham:bfklm}); (b) The schematic phase diagram of BFKLM, showing a continuous quantum phase transition between two phases with large and small Fermi surfaces, respectively. The magenta line is the phase boundary, specifying the manifold of QCPs.
 }
\label{fig:model}
\end{figure}

\sect{Bose-Fermi Kondo lattice model and its dynamical large-$N$ limit} 
We introduce here the BFKLM, 
which is illustrated in Fig.\,\ref{fig:model}(a). For
our purpose,
{the spin symmetry is generalized from SU(2) to} 
$SU(N)$.
The Hamiltonian reads
\begin{align}
\label{ham:bfklm}
    H= &H_{\mathrm{K}} + H_{\mathrm{B}} \\
    H_{\mathrm{K}}=&\sum_{\bm{k}\sigma} \epsilon_{\bm{k}}c^{\dagger}_{\bm{k}\sigma}c_{\bm{k}\sigma}+\frac{J_\mathrm{K}}{N}\sum_{i\sigma\sigma'}\boldsymbol{S}_{i}\cdot(c^{\dagger}_{i\sigma}\boldsymbol{\tau}_{\sigma\sigma'}c_{i\sigma'})\nonumber
    \\H_{\mathrm{B}}=&\sum_{\bm{q}}w_{\bm{q}}\boldsymbol{\Phi}^{\dagger}_{\bm{q}}\cdot\boldsymbol{\Phi}_{\bm{q}}+\frac{g}{\sqrt{N}}\sum_{i}\boldsymbol{S}_{i}\cdot(\boldsymbol{\Phi}_{i}+\boldsymbol{\Phi}_{i}^{\dagger}).\nonumber
\end{align}
Here, 
the local moments ($\mathbf S_i$)  interact with conduction electrons 
($c_{i \sigma}$) via a Kondo coupling $J_\mathrm{K}$, and with vector bosons ($\mathbf \Phi_i$) via a bosonic coupling $g$.
The local moment spin operator \( \bm{S}_i \) transforms under the adjoint representation of the global \( SU(N) \) spin symmetry and is represented by $f$-electrons:
$
S_{i\sigma\sigma'} = \bm{S}_{i}\cdot\bm{\tau}_{\sigma'\sigma}=f^\dagger_{i\sigma} f_{i\sigma'} - \frac{1}{2} \delta_{\sigma\sigma'}
$, {where $\bm \tau$ describes the $N^2 - 1$ generators of $SU(N)$  in the fundamental representation}. Furthermore,
\( c_{i\sigma} \) denotes the conduction electron at site \( i \) with spin index \( \sigma = 1, \dots, N \), and \( \epsilon_{\mathbf{k}} \) is the band dispersion. 
It is important to note that both the local moment spins and $c$-electrons have an $SU(N)$ symmetry; in other words,
there is exactly one channel ($M=1$) for each spin species and, thus, $\kappa \equiv M/N = 1/N$. This ensures that 
the Kondo effect is of the single-channel exactly screeened (as opposed to multi-channel over-screened) {type.}
The vector bosons \( \mathbf{\Phi}_i \), also in the adjoint representation, have \( N^2 - 1 \) components,
are associated with the collective spin fluctuations and have an energy dispersion \( w_{\mathbf{q}} \).
Their introduction here is inspired by the EDMFT approach~\cite{Hu-Natphys2024,lcp-prb}.
The \( SU(N) \) generators \( \boldsymbol{\tau}_{\sigma\sigma'} \) act on the conduction electron spin indices. 
The exponent $\gamma$ controls the energy scaling of the
bare local 
spectral function of the $\mathbf \Phi$ bosons:
$A_{\Phi}(\omega)=\int_{\bm{q}}A_{\Phi}(\omega,\bm{q})=2\pi K_{0}^{2}|\omega|^{\gamma}\sign(\omega)$, where $A_{\Phi}(\omega,\bm{q})=i(G^{R}_{\Phi}(\omega,\bm{q})-G^{A}_{\Phi}(\omega,\bm{q}))=\pi[\delta(\omega-w_{\bm{q}})-\delta(\omega+w_{\bm{q}})]$ is the bare spectral function for the vector bosons;
while $\gamma$ is an input value in our model, to capture the EDMFT
results~\cite{Hu-Natphys2024,Si-Nature,lcp-prb}, we will focus on the case of $\gamma = 0^{+}$ (see the SM).
Note that the BFKLM represents
a lattice generalization of the   Bose-Fermi Kondo 
model~\cite{Si.96,SmithSi_EPL1999,Sengupta,Zhu2002,ZarandDemler,Zhu2004},
with the Kondo coupling $J_{\mathrm{K}}$ and bosonic coupling $g$ 
being extended to each site of the lattice.
{The Bose-Fermi Kondo model~\cite{Si.96,SmithSi_EPL1999,Sengupta,Zhu2002,ZarandDemler,Zhu2004}
{\it per se} exemplifies the type of effective
models that arise in the EDMFT approach to underlying correlated electron lattice Hamiltonians~\cite{Hu-Natphys2024,Si.96,SmithSi-edmft,Chitra2001}.}

We stress that the first part ($H_\mathrm{K}$) on its own is the venerable Kondo lattice Hamiltonian.
In the standard large-$N$ approach to
$H_{\mathrm{K}}$,
RKKY interactions mediated by the conduction electrons is suppressed relative to the Kondo effect as $N$ becomes large~\cite{Hewson}; 
indeed, the RKKY and related spin-exchange interactions only appear
at order $1/N^2$ and beyond (Refs.~\onlinecite{Si-Lu-Levin1992,Houghton-Read-Won1988}).
By contrast, in the BFKLM, the coupling of the local moments with the vector bosons from $H_{\mathrm{B}}$ 
promotes an effective RKKY interaction to the leading order in $1/N$.
Thus, the 
bosonic coupling introduces the dynamical effect of the RKKY interaction.
The resulting interplay between the Kondo and bosonic couplings
will be shown to yield
a Kondo destruction QCP.

To perform the large-$N$ analysis,
we decouple the Kondo coupling term $
(J_\mathrm{K}/N)
\boldsymbol{S}_{i}\cdot\sum_{\sigma\sigma'}(c^{\dagger}_{i\sigma}\boldsymbol{\tau}_{\sigma\sigma'}c_{i\sigma'})$ into
\begin{align}
   \frac{1}{\sqrt{N}}\bar{B}_{i}\sum_{\sigma}c^{\dagger}_{i\sigma}f_{i\sigma}+h.c+\frac{\bar{B}_{i}B_{i}}{
   J_\mathrm{K}}
\end{align}
 via a Hubbard–Stratonovich transformation. Here, $\bar{B}_{i}$ is the hybridization field, which carries the same electromagnetic $U(1)$ charge as the conduction electron $c_{i\sigma}$, while the $f$-electrons are charge-neutral. 

\sect{Phase diagram and Kondo destruction QCP}
For $g=0$, the $B$-field is condensed and $B_{i}(t)=\sqrt{N}b$ where $b$ is the static hybridization field  
that is determined self-consistently~\cite{Hewson}.
We note that $b$ captures the strength
of Kondo screening.
The ground state is the Kondo screening phase where the 
$f$- and conduction $c$- electrons
hybridize to form a large Fermi surface. 
Accompanying this formation of the large Fermi surface is the transfer 
of the $c$-electron's 
U(1) charge to $f$-electrons, which partake in the formation 
of the heavy quasiparticles.

When $\gamma<1$, and this includes the case of our focus, $\gamma=0^+$, we find that
the bosonic coupling $g$ suppresses the Kondo screening at $g=g_c$ such that $b=0$  and induces a quantum phase transition that destroys the Kondo effect
(see the next section).
For $g > g_c$, the  $B$-field is uncondensed  
and the Fermi surface is formed by 
 the conduction electrons alone -- this is the small Fermi surface. 
 
The resulting phase diagram is illustrated in Fig.\,\ref{fig:model}(b).
 Thus, our model captures the key 
 physics of Fermi surface jump across the Kondo destruction QCP.
 
\begin{figure}
    \centering
    \includegraphics[width=\linewidth]{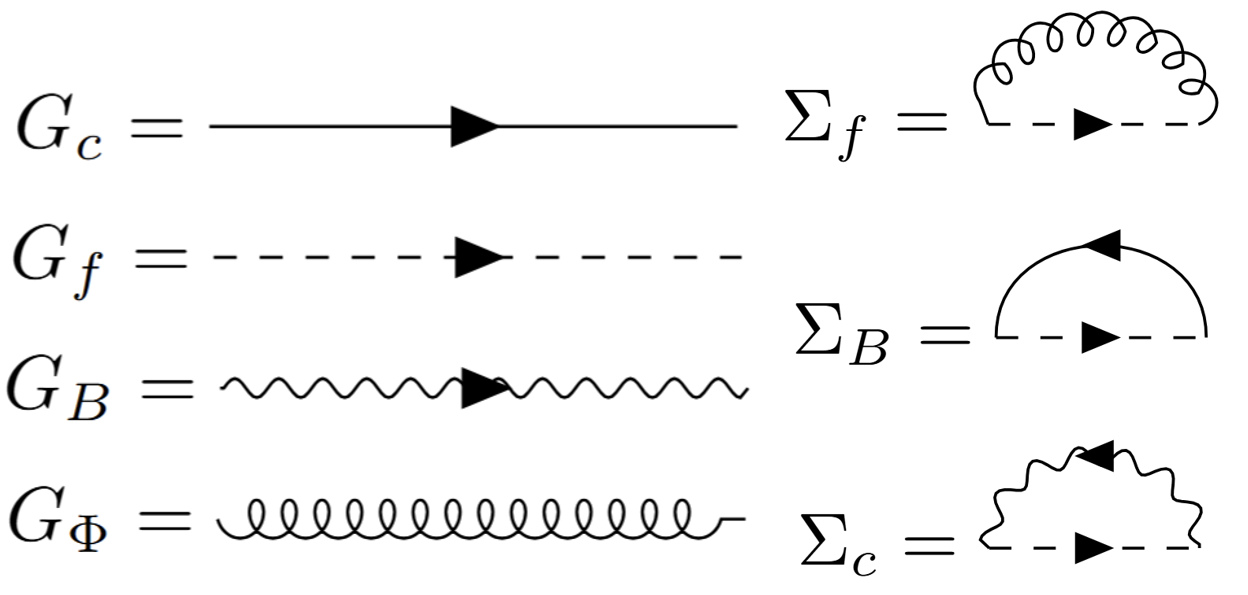}
    \caption{Feynman rules and large-$N$ self-energy diagrams in the Bose-Fermi Kondo lattice model.
        Solid lines denote conduction electron propagators \( G_c \), dashed lines represent $f$-electron propagators \( G_f \),
        wavy lines indicate hybridization propagator \( G_B \), and curly lines correspond to the vector boson propagator \( G_\Phi \).
        The diagram for \( \Sigma_f \) represents the $f$-electron self-energy due to a coupling with \( G_\Phi \),
        and the diagram for \( \Sigma_B \) shows the bosonic self-energy due to the fermionic fluctuations. The diagram for $\Sigma_{c}$ represents the conduction electron self-energy from the Kondo coupling.}
    \label{fig:diagram}
\end{figure}

 \sect{Keldysh formulation and equilibrium properties} 
 We now study the system at the 
 QCP using the Keldysh formalism, a Green's function approach that captures both equilibrium and nonequilibrium dynamics \cite{haug2008quantum,kamenev2023field}. 
 In the large-$N$ limit, the model yields the following saddle-point equations defined on the Keldysh contour $C$:
\begin{align}    
    \Sigma_{f}(x_1,x_2)=&\,ig^{2}G_{\Phi}(x_1,x_2)G_{f}(x_1,x_2)\label{fself},\\
    \Sigma_{B}(x_1,x_2)=&-iG_{c}(x_2,x_1)G_{f}(x_1,x_2) ,\label{bself}
\end{align}
where $x_{1/2}=(\bm{r}_{1/2},t_{1/2})$ is the combined notation of site $\bm{r}$ and time $t$. $\Sigma_{f/B}(x,x')$ is the self-energy of $f$-electron/ dynamical hybridization field $B$ defined via: $G(x_1,x_2)=G_{0}(x_{1},x_2)+\int_{C}{dx_{3}dx_{4}}G_{0}(x_1,x_3)\Sigma(x_3,x_4)G(x_4,x_2)$. The contour-ordered Green's functions are defined as follows, $G_{c}(x,x')=-i\langle T_{C}c_{i\sigma}(t)c_{j\sigma}^{\dagger}(t') \rangle$,  $G_{f}(x,x')=-i\langle T_{C}f_{i\sigma}(t)f_{j\sigma}^{\dagger}(t')\rangle$, $G_{B}(x,x')=-i\langle T_{C}B_{i}(t)B_{j}^{\dagger}(t')\rangle$, $G_{\Phi}(x,x')=-i\langle T_{C}(\Phi_{i}+\Phi_{i}^{\dagger})(t)(\Phi_{j}+\Phi_{j}^{\dagger})(t')\rangle$, where $T_{C}$ is the contour-time ordering operator. The diagrammatic representation is shown in Fig.\,{\ref{fig:diagram}}. 
The conduction electrons'   contribution to the  self-energy of the local $f$-electrons    is $O(1/N)$, 
given that there is only a single channel 
of conduction electrons (for each spin component).
The self-energies of the conduction electrons and vector bosons are also of the order $O(1/N)$:
\begin{align}
    \Sigma_{c}(x_1,x_2)&=\frac{i}{N}G_{B}(x_2,x_1)G_{f}(x_1,x_2)\label{cself},\\
    \Sigma_{\Phi}(x_1,x_2)&=-\frac{ig^2}{N}G_{f}(x_2,x_1)G_{f}(x_1,x_2),\label{phiself}
\end{align}

At equilibrium (without an applied voltage bias), the system is space-time translationally invariant: $G(x,x')=G(x-x')$. 
We determine the quantum critical behavior by solving the equations with the scale-invariant ansatz for the spectral functions;
the details of the procedure are given in the SM~\cite{sm}. We obtain the solutions for the Green's functions $G_f$ and $G_B$ at the Kondo destruction QCP:
\begin{align}
G_{f}^{R}(\omega)=&\frac{\sign(\omega)}
{\sqrt{2}K_{0}g_{c}}
(i\omega)^{-\frac{1}{2}} \, , 
\\G_{B}^{R}(\omega)=&-\frac{K_{0}g_{c}}{\sqrt{2}\rho }
(-i\omega)^{-\frac{1}{2}} \, ,
\label{GB}
\end{align}
where $\rho$ is the density of states for conduction electrons at the Fermi level. The spin susceptibility on the Keldysh contour is defined as
    $\chi(x_{1}, x_{2}) = -\frac{i}{N^{2}} \sum_{\sigma\sigma'} \left\langle T_{C}f^{\dagger}_{i\sigma}(t) f_{i\sigma'}(t) f^{\dagger}_{j\sigma'}(t') f_{j\sigma}(t') \right\rangle $.
    To the leading order in $1/N$, $\chi(x_{1}, x_{2}) = -i\, G_{f}(x_{2}, x_{1}) G_{f}(x_{1}, x_{2}) $.
    At this order, it is local: 
    $\chi^{R}(\bm{q}, \omega) = \chi^{R}_{\text{loc}}(\omega)$; wavevector-dependent terms appear in the next order in $1/N$  (see the SM). To the leading order in $1/N$,
the local dynamical spin susceptibility is
the convolution of $G_f(\omega)$.
The result  
is regularized by a ultraviolet (UV) cut-off energy scale, $\omega_c$, and is given by:
\begin{align}
    \chi_{loc}^{R}
    (\omega)
    =
        -\frac{1}{2\pi(K_{0}g_{c})^2} \ln\left(\frac{\omega_c}{-i\omega} \right)
    \, .
\end{align}
Our large-$N$ result is consistent with the EDMFT result [{\it cf.}, Eq.\,(56) of Ref.\,\onlinecite{lcp-prb}].

As in the EDMFT solution, our results indicate that the Kondo destruction QCP
is controlled by an interacting fixed point, which in turn implies that the dynamics 
satisfy 
the dynamical ($\hbar \omega /k_{\rm B} T$) scaling at nonzero temperatures.

\sect{Quantum kinetic equations}
 Having established the existence and equilibrium signatures of the Kondo destruction QCP in the large-$N$ limit of the BFKLM, we now turn to its nonequilibrium properties 
 in current-carrying steady states under an applied electric field $\bm{E}$. We introduce a voltage bias across the system $V_{\text{field}}=-\sum_{i\sigma}e\bm{E}\cdot\bm{r}_{i}c^{\dagger}_{i\sigma}c_{i\sigma}$,  and include a disorder potential $H_{\text{disorder}}=\sum_{i\sigma}u_{i}c^{\dagger}_{i\sigma}c_{i\sigma}$ to drive diffusive transport, where $u_{i}$ is a random disorder potential whose disorder average $\overline{u_{i}}=0,\, \overline{u_{i}u_{j}}=u^2\delta_{i,j}$. The combination of the bias and disorder gives rise to a spatial profile of the electronic distribution functions.

We derive quantum kinetic equations using the Keldysh formalism in the presence of disorder (see the SM~\cite{sm}). A key feature of the large-$N$ solution is that only the conduction electrons and the hybridization \( B \) field are directly affected by the voltage bias at the leading order in $1/N$. In contrast, the $f$-electrons and the vector bosons $\Phi$, being charge-neutral, remain in equilibrium to the leading order.

The distribution function \( n_c(\bm{k},\epsilon,x) \) of the conduction electrons satisfies a semiclassical Boltzmann equation. Decomposing it into even and odd components in the momentum space, we find that the even part obeys a diffusion equation,
\begin{align}
    D \partial_x^2 n_{c}(\epsilon, x) = I_{ee}^{\mathrm{coll}}(\epsilon, x) + I_{B}^{\mathrm{coll}}(\epsilon, x),
\end{align}
where \( D \) is the diffusion constant. $I_{ee}^{\mathrm{coll}}(\epsilon,x)$ corresponds to the electron-electron collision integral introduced for completeness, and \( I_{B}^{\mathrm{coll}}(\epsilon, x) \) encodes inelastic scattering of the conduction electrons mediated by the hybridization field $B$. In regimes of strong inelastic scattering, the electron distribution relaxes to a local Fermi-Dirac form with position-dependent temperature \( T_e(x) \) and chemical potential \( \mu(x) = -eEx \).

\begin{figure}
\centering
\includegraphics[width=.75\columnwidth]{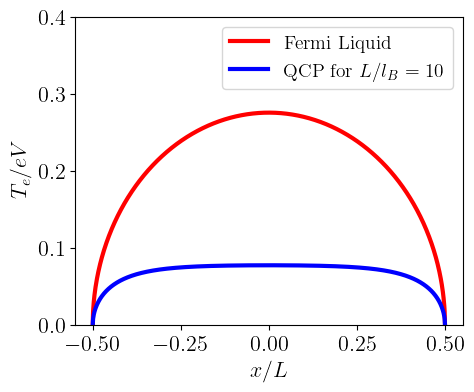}%
\caption{Spatial profile of the local electronic temperature \( T_{e}(x) \) for a Fermi liquid and at the Kondo destruction 
QCP, shown for \( L / l_{B} = 10 \), where
  $l_{\rm B}$ represents the relaxation length 
  of 
  the conduction electrons from the critical Kondo interaction processes
  mediated by the nonequilibrium dynamical $B$ field.  
}
    \label{fig:temp}
\end{figure}

Solving the diffusion equation, as detailed in the SM, leads to the spatial profile of the local electronic temperature:
\begin{align}
    T_e(x) = \frac{\sqrt{6}\,(eV) l_B}{\pi L} \sqrt{1 - \frac{\cosh(x/l_B)}{\cosh(L/2l_B)}} \, , 
\end{align}
where \(V=EL\) is the applied voltage bias across the system of size $L$, and 
\( l_B \)
is the
relaxation length of 
the 
conduction electrons from the critical Kondo interaction process that involves 
the 
nonequilibrium dynamical $B$-field. This temperature profile, shown in Fig.~\ref{fig:temp}, demonstrates how energy dissipates in the presence of the Kondo-destruction quantum criticality.

\sect{Shot noise}
We study the shot noise:
\begin{align}
    S=2\int dt \langle \delta I(t) \delta I(0) \rangle
\end{align}
where $\delta I(t)=I(t)-\langle I\rangle$ is the current fluctuation around the average current $\langle I \rangle$.
To proceed, we express the noise in terms of the Green's functions by employing the $1/N$ expansion,
\begin{align}
    S&=2e^2\frac{N}{L}\int_{-L/2}^{L/2} dx \sum_{k}v_{k_x}^{2}\int \frac{d\omega}{2\pi}G_{c}^{>}(k,\omega,x)G_{c}^{<}(k,\omega,x)\nonumber\\
    &=2e^2N_{L}\frac{v_{F}^2\rho}{4d}\frac{N}{L}\int _{-L/2}^{L/2}dx \int d\omega\frac{n_{c}(\omega,x)[1-n_{c}(\omega,x)]}{|\mathrm{Im}\Sigma_{c}(\omega)|},
\end{align}
where $v_{k_{x}}=\partial\epsilon_{k}/\partial k_{x}$ is the electron group velocity along the transport direction, approximated as $v_{F}/\sqrt{d}$, with $v_{F}$ the Fermi velocity.
In the zero temperature limit and to the leading order of $1/N$, the impurity scattering dominates over the inelastic scatterings and, thus, $|\mathrm{Im}\Sigma_{c}(\omega)|=\frac{1}{\tau}$. Accordingly,
\begin{align}
    S=4G\bar{T}_{e}
\end{align}
where $\bar{T}_{e}=\frac{1}{L}\int_{-L/2}^{L/2} dx T_{e}(x)$ is the averaged local temperature, 
and $G=e^{2}NN_{L}\rho v_{F}^{2}\tau/8d$ is the conductance at zero temperature.  

\begin{figure}
\centering
\includegraphics[width=.75\columnwidth]{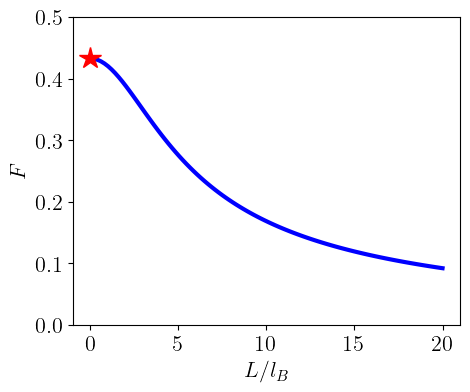}%
\caption{
  The shot noise Fano factor $F$ as a function of the normalized system size \( L / l_{\mathrm{B}} \) at the Kondo destruction QCP, where 
  $l_{\rm B}$ represents the
 relaxation
  length of the conduction electrons 
 from the critical Kondo interaction processes mediated by the nonequilibrium dynamical $B$ field. The red star marks the Fermi liquid value $F=\sqrt{3}/4$.
}
\label{fig:fano}
\end{figure}

The Fano factor, defined as the ratio of the current noise to the average current, has the following analytic form at the 
Kondo destruction QCP:
\begin{align}
    F=&\frac{S}{2GeV} \nonumber \\
    =&-i\frac{8\sqrt{6}}{\pi \eta^2}E\left(\frac{i\eta}{4},i\csch\frac{\eta}{4}\right)\sqrt{\tanh\frac{\eta}{4}\tanh\frac{\eta}{2}}\label{eq:fano}
\end{align}
where $\eta=L/l_{B}$, $E(\phi,k)=\int_{0}^{\phi}d\theta\sqrt{1-k^{2}\sin^{2}\theta}$ denotes the incomplete elliptic integral of the second kind. 
Fig.\,\ref{fig:fano} shows the Fano factor as a function of $L/l_{B}$.
The result shows that the Fano factor is strongly suppressed below the Fermi liquid ($F=\sqrt{3}/4$) and Fermi gas ($F=1/3$) value
when $L/l_{B} \gtrsim 5$, and decays as $1/L$ in the infinite system size limit.

\sect{Discussion and conclusion}
{Our finding of} a pronounced suppression of the Fano factor with increasing system size
{is consistent with the experimental results} 
in YbRh\(_2\)Si\(_2\)~\cite{Natelson2022}.
{In the experiments, the}
observed reduction 
was shown to {be due to electron-electron interactions, given that it}
significantly exceeds what can be accounted for by the electron-phonon coupling alone~\cite{Natelson2022}.

Theoretically, 
in our approach, the vector $\boldsymbol{\Phi}$ boson couples only to the charge-neutral degrees of freedom. In this sense, they are shielded from the current-carrying degrees of freedom. This is reflected in the lack of any effect 
by the applied electric field to drive the $\boldsymbol{\Phi}$ boson out of equilibrium up to order $1/N$. By contrast, 
the hybridization boson $B$ couples directly to the current-carrying $c$-electrons and, accordingly, it acquires a nonequilibrium distribution in the presence of an applied electric field. Our results show that the reduction of the 
Fano factor appears through the 
{relaxation} of the current-carrying $c$-electrons 
by the processes that involve the hybridization $B$ boson.

To put it more broadly, the reduction of the Fano factor we have found 
reflects the loss of quasiparticles in the Kondo destruction quantum critical regime. 
If the heavy quasiparticles had developed, the hybridization $B$ field would have
condensed; in that case,
the $B$ boson would cease to exist as a low-energy scatterer, and the shot noise reduction mechanism we have advanced would no longer operate.

To summarize, we have theoretically studied current fluctuations in 
{a model for heavy fermion strange metals. We have done so by advancing}
a minimal Bose-Fermi Kondo lattice model.
{The model} allows for a well-defined large-$N$ limit and,
{in equilibrium,} captures
salient behavior of the
Kondo destruction quantum criticality.
We found a strong suppression of the shot noise in the quantum critical regime, thereby providing the understanding of the striking recent experiment in YbRh$_2$Si$_2$. 
 By linking it to the loss of quasiparticles in a well-defined theoretical model,
 our work points to the shot noise reduction as a means to directly probe the nature of charge carriers in strange metals.
 {Given the expanding horizons of strange metallicity
 and the broad implications of its physics~\cite{Savitsky2025},
 our work raises the prospects for deepened understanding of correlated quantum materials in general and the overarching electron localization-delocalization phenomena in particular.}

\paragraph*{{\bf Acknowledgements: }}
We thank
{Matt Foster, Chandan Setty and Tsz Chun Wu} 
for useful discussions.
This work has been supported in part by the NSF Grant No.\ DMR-2220603 (Y.W.,F.X.), the AFOSR under 
Grant No.\ FA9550-21-1-0356 (S.S.,H.H.), the Robert A. Welch Foundation Grant No.\ C-1411 (Q.S.) and 
the Vannevar Bush Faculty Fellowship ONR-VB N00014-23-1-2870 (Q.S.).
The majority of the computational calculations have been performed on the Shared University Grid at Rice funded by NSF under Grant No.~EIA-0216467, a partnership between Rice University, Sun Microsystems, and Sigma Solutions, Inc., the Big-Data Private-Cloud Research Cyberinfrastructure MRI-award funded by NSF under Grant No. CNS-1338099, and the Advanced Cyberinfrastructure Coordination Ecosystem: Services \& Support (ACCESS) by NSF under Grant No. DMR170109.
{The work of D.N. has in part been supported by the DOE BES under Award No.\ DE-FG02-06ER46337.
S.P. acknowledge funding by the European Union (ERC Adv. Grant CorMeTop, project 101055088), the Austrian Science Fund (FWF) through
the projects SFB F 86 (Q-M\&S), FOR 5249 (QUAST), and 10.55776/COE1 (quantA), and the US AFOSR through project FA8655-24-1-7018 (CorTopS).}
Q.S. acknowledges the hospitality of the Aspen Center for Physics, which is supported by NSF grant No. PHY-2210452.

\bibliography{shotnoise}
\bibliographystyle{apsrev4-2}

\clearpage
\onecolumngrid

\setcounter{secnumdepth}{3}

\onecolumngrid
\newpage
\beginsupplement

\section*{Supplemental Materials}

\tableofcontents

 \section{
 Pertinent Kramers-Kronig relations}\label{Sec:A}
In this section we 
enumerate the useful Kramers-Kronig relations for solving the large-$N$ equations in the main text.
For a fermionic spectral function which exhibit the
scale-invariant form,
\begin{align}
    A(\omega)=|\omega|^{-\alpha} \, ,
\end{align}
where $0<\alpha<1$,
the corresponding retarded Green's function has the following form:
\begin{align}
G^{R}(\omega)&=
    \int\frac{d\omega'}{2\pi}\frac{A(\omega')}{\omega-\omega'+i0^{+}}\\
    &=-i\int\frac{d\omega'}{2\pi}|\omega'|^{-\alpha}\int_{0}^{\infty} d\lambda \, e^{i\lambda(\omega-\omega'+i0^+)}\\
    &=-i\int_{0}^{\infty} d\lambda \, e^{i\lambda(\omega+i0^+)}\int\frac{d\omega'}{2\pi}|\omega'|^{-\alpha}e^{-i\lambda\omega'}\\
    &=-i\int_{0}^{\infty} d\lambda \, e^{i\lambda(\omega+i0^+)}\frac{\sin(\frac{\pi\alpha}{2})\Gamma(1-\alpha)}{\pi|\lambda|^{1-\alpha}}\\
    &=-i\frac{\sin(\frac{\pi\alpha}{2})\Gamma(1-\alpha)}{2\pi}\left[\mathrm{Re}\int_{-\infty}^{\infty}d\lambda\frac{e^{i\lambda\omega}}{|\lambda|^{1-\alpha}}+i\,\mathrm{Im}\int_{-\infty}^{\infty}d\lambda\frac{e^{i\lambda\omega}\sign(\lambda)}{|\lambda|^{1-\alpha}}\right]\\
    &=\frac{\sign(\omega)}{2\sin{\left[\frac{\pi}{2}(1-\alpha)\right]}}|\omega|^{-\alpha}e^{-i\frac{\pi}{2}(1-\alpha)\sign(\omega)} \, .
\end{align}
Specifically for $\alpha=0$, we have:
\begin{align}
\int\frac{d\omega'}{2\pi}\frac{1}{\omega-\omega'+i0^{+}}&=-\frac{i}{2} .
\end{align}
Similarly, for its bosonic
counterpart,
\begin{align}
    A(\omega)=|\omega|^{-\alpha}\sign(\omega) \, ,
\end{align}
the retarded Green's function has the form:
\begin{align}
    G^{R}(\omega)&=\int\frac{d\omega'}{2\pi}\frac{A(\omega')}{\omega-\omega'+i0^{+}}\\&= 
    -\frac{1}{2\sin{\left[\frac{\pi\alpha}{2}\right]}}|\omega|^{-\alpha}e^{i\frac{\pi\alpha}{2}\sign(\omega)},
\end{align}

Accordingly,
the local Green's function for the conduction electrons has the following form,
\begin{align}
G_{c}^{R}(\omega)=\int_{k}G_{c}^{R}(\bm{k},\omega)=\int d\epsilon \frac{\rho(\epsilon)}{\omega-\epsilon+i0^+}\approx-i\pi\rho \, ,
\end{align}
where $\rho(\epsilon)=\int_{k}\delta(\epsilon-\epsilon_{\bm{k}})\approx \rho$ is the density of states approximated at the Fermi level.

Likewise, for
the vector bosons, the bare local spectral function is 
taken to be sub-Ohmic (see the main text): $A_{\Phi}(\omega)=\int_{q}\delta(\omega-w_{p})-\delta(\omega+w_{p})=2\pi K_{0}^{2}|\omega|^{\gamma}\sign(\omega)$.
In the SM, we will not only consider the case of our focus, $\gamma=0^{+}$, but also the more general 
cases with $0<\gamma<1$. The corresponding 
retarted Green's function is:
\begin{align}
    G_{\Phi}^{R}(\omega)=G_{\Phi}^{R}(0)+\omega\int\frac{d\omega'}{2\pi}\frac{2\pi K_{0}^{2}|\omega'|^{\gamma-1}}{\omega-\omega'+i0^{+}}=G_{\Phi}^{R}(0)+\frac{\pi K_{0}^{2}}{\sin{\left(\frac{\pi\gamma}{2}\right)}}|\omega|^{\gamma}e^{-i\frac{\pi\gamma}{2}\sign(\omega)}.
\end{align}

\section{Equilibrium solution}
{In the absence of an applied voltage,
space-time translational invariance dictates that the following:} $G(x,x')=G(x-x')$. 
{When the Kondo effect is destroyed,} we find that $G_{f}$ is local. 
{We} can appreciate the feature 
from the following considerations. When the Kondo effect is destroyed,
the fermion occupation number $\sum_{\sigma}f^{\dagger}_{i\sigma}f_{i\sigma}=N/2$ is 
a good quantum number at every site. Accordingly, 
 the Green's function $G_{f}^{R}(\bm{r}_{i}-\bm{r}_{j},t-t')=-i\langle \{f_{i\sigma}(t),f^{\dagger}_{j\sigma}(t')\}\rangle\theta(t-t')  \propto \delta_{i,j}$ 
 vanishes for $i\neq j$. 
{Thus,}
$G_{f}(\bm{k},\omega) = G_{f}(\omega)$.

{Our focus is for the QCP. With the locality, we can take 
the following form,}
$A_{f}(\omega)=2C|\omega|^{-\alpha_{f}}$, with $0<\alpha_{f}<1$, 
{for the $f$-electron spectral function,
$A(\omega)=i[G^{R}(\omega)-G^{A}(\omega)]$.} 
The retarded and advanced Green's functions are 
obtained using the Kramers-Kronig relations (see the above section in the SM).
In addition, the hybridization field $B$ is critical at the
QCP, with $1/J + \Sigma_B(0) = 0$. 

By solving the equations with the
{scale-invariant} 
ansatz for the spectral function, we obtain the solutions for the Green's functions $G_f$ and $G_B$ at the Kondo destruction QCP, with $\alpha_{f}=1/2$:
\begin{align}
G_{f}^{R}(\omega)=&\frac{\sign(\omega)}{\sqrt{2}K_{0}g_{c}}e^{-i\frac{\pi}{4}\sign(\omega)}|\omega|^{-\frac{1}{2}},\\G_{B}^{R}(\omega)=&-\frac{K_{0}g_{c}}{\sqrt{2}\rho }e^{i\frac{\pi}{4}\sign(\omega)}|\omega|^{-\frac{1}{2}},\label{GB}
\end{align}
where $\rho$ is the density of states for conduction electrons at the Fermi level.

{The spin susceptibility on the Keldysh contour is defined as
\begin{align}
    \chi(x_{1}, x_{2}) = -\frac{i}{N^{2}} \sum_{\sigma\sigma'} \left\langle f^{\dagger}_{i\sigma}(t) f_{i\sigma'}(t) f^{\dagger}_{j\sigma'}(t') f_{j\sigma}(t') \right\rangle = -i\, G_{f}(x_{2}, x_{1}) G_{f}(x_{1}, x_{2}) \,.
\end{align}
{The last equality is valid to the leading order in \(1/N\). To this order,
$\chi^{R}(\bm{q}, \omega) = \chi^{R}_{\text{loc}}(\omega)$\,.}
}
The local dynamical spin susceptibility is, to the leading order in $1/N$, the convolution of $G_f(\omega)$.
The result 
is regularized by a ultraviolet (UV) cut-off energy scale, $\omega_c$, and is given by:
\begin{align}
    \chi_{loc}^{R}(\omega)=&\frac{1}{2(K_{0}g_{c})^2}\int \frac{d\omega'}{2\pi}\frac{\sign{(\omega')}}{\omega-\omega'+i0^{+}}\\
    =&-\frac{1}{2\pi(K_{0}g_{c})^2}
    \ln\left(\frac{\omega_c}{-i\omega} \right)
\end{align}
Our large-$N$ result is consistent with the EDMFT result ({\it cf.}, Eq.\,(56) of Ref.\,\onlinecite{lcp-prb}).

The result above applies to leading order in $1/N$. 
Corrections at the next order will introduce momentum dependence,
with the inverse of the lattice dynamical spin susceptibility
now experiencing 
a correction of $- \frac{g_{c}^{2}}{N} G_{\Phi}^{R}(\bm{q}, \omega)$
on top of the inverse local spin susceptibility.

So far we have considered the solution of the large-$N$ equations for $\gamma=0^{+}$. [As mentioned in the main text, this was specified based on the self-consistent solution determined in the EDMFT approach, {\it cf.} Eq.\,(54) of Ref.\,\onlinecite{lcp-prb}.]
For completeness, we
describe the solutions for general $\gamma$ in the range $0<\gamma<1$, for which a QCP is realized. The corresponding solutions at the QCP are as follows:
\begin{align}
G_{f}^{R}(\omega)=&\frac{C\sign(\omega)}{\sin\left(\frac{\pi}{2}\frac{1-\gamma}{2}\right)}e^{-i\frac{\pi}{2}\frac{1-\gamma}{2}\sign(\omega)}|\omega|^{-\frac{1+\gamma}{2}},\\G_{B}^{R}(\omega)=&-\frac{(1-\gamma)\sin\left(\frac{\pi}{2}\frac{1-\gamma}{2}\right)}{2\rho C}e^{i\frac{\pi}{2}\frac{1-\gamma}{2}\sign(\omega)}|\omega|^{-\frac{1-\gamma}{2}}\label{GB1}
\end{align}
where $
C=\frac{\sin\left(\frac{\pi}{2}\frac{1-\gamma}{2}\right)}{K_{0}g\sqrt{B\left(1+\gamma,\frac{1-\gamma}{2}\right)}}    
$; here, $B(x,y)$ is the beta function and $\rho$ is the density of states for the conduction electrons at the Fermi level. 

The dynamical local spin susceptibility is, to the leading order in $1/N$, the convolution of $G_f(\omega)$.
The result for the retarted form,
$\chi_{\text{loc}}^{R}(\omega)$, is as follows:
\begin{align}
    \chi_{\text{loc}}^{R}(\omega)
    = -\frac{C^{2} \Gamma\left(\frac{1 - \gamma}{2}\right)^2}{\pi \Gamma(1 - \gamma)\sin{\frac{\pi\gamma}{2}}} 
    |\omega|^{-\gamma} e^{i\frac{\pi\gamma}{2} \, \text{sgn}(\omega)}.
\end{align}

For completeness, we also write down these Green functions and susceptibilities in Masubara frequencies, 
which are obtained from the corresponding spectral functions through the Kramers-Krönig relations:
\begin{align}
    G(i\omega_{n})=\int\frac{d\omega}{2\pi}\frac{A(\omega)}{i\omega_{n}-\omega}
\end{align}
where $A(\omega)=-2\mathrm{Im}G^{R}(\omega)$.
The corresponding results are as follows: 
\begin{align}
    G_{f}(i\omega_{n})&=\frac{-iC}{\sin\left(\frac{\pi}{2}\frac{1-\gamma}{2}\right)}|\omega_{n}|^{-\frac{1+\gamma}{2}}\\
    G_{B}(i\omega_{n})&=-\frac{(1-\gamma)\sin\left(\frac{\pi}{2}\frac{1-\gamma}{2}\right)}{2\rho C}|\omega_{n}|^{\frac{1-\gamma}{2}}\\
    \chi_{loc}(i\omega_{n})&=-\frac{C^{2} \Gamma\left(\frac{1 - \gamma}{2}\right)^2}{\pi \Gamma(1 - \gamma)\sin{\frac{\pi\gamma}{2}}} 
    |\omega_{n}|^{-\gamma} 
\end{align}
At $\gamma=0^{+}$,
the corresponding 
result is:
\begin{align}
    \chi_{loc}(i\omega_{n})=&-\frac{1}{(K_{0}g)^2}\int_{0}^{\omega_{c}}\frac{d\omega}{2\pi}\frac{\omega}{\omega^2+\omega_{n}^{2}}\\
 =&-\frac{1}{2\pi(K_{0}g_{c})^2}\ln\left(\frac{\omega_{c}}{|\omega_{n}|}\right)   
\end{align}

Green's functions in the imaginary time are determined by:
\begin{align}
    G(\tau)=\int\frac{d\omega}{2\pi}\frac{-e^{-\tau\omega}}{1\pm e^{-\beta\omega}}A(\omega)
\end{align}
where the $+/-$ sign corresponds to fermion/boson.
We have 
\begin{align}
    G_{f}(\tau)=&
\frac{ \, C\beta^{-\frac{1-\gamma}{2}} \, \Gamma\left(\frac{1 - \gamma}{2}\right)}{2^{\frac{1-\gamma}{2}}\pi} \left(
  -\zeta\left(\frac{1 - \gamma}{2}, \frac{\beta - \tau}{2\beta}\right)
  - \zeta\left(\frac{1 - \gamma}{2}, \frac{\tau}{2\beta}\right)
  + \zeta\left(\frac{1 - \gamma}{2}, \frac{\beta + \tau}{2\beta}\right)
  + \zeta\left(\frac{1 - \gamma}{2}, 1 - \frac{\tau}{2\beta}\right)
\right)\\
= &-\frac{1}{\pi}C\Gamma\left(\frac{1-\gamma}{2}\right)\left(\frac{1}{\tau}\right)^{\frac{1-\gamma}{2}} \text{for $\tau/\beta\ll 1$}\\
    G_{B}(\tau)=&-\frac{(1-\gamma)\sin^{2}\left(\frac{\pi}{2}\frac{1-\gamma}{2}\right)}{\rho C}\beta^{-\frac{1+\gamma}{2}}\Gamma\left(\frac{1+\gamma}{2}\right)\left[\zeta\left(\frac{1+\gamma}{2},1-\frac{\tau}{\beta}\right)+\zeta\left(\frac{1+\gamma}{2},\frac{\tau}{\beta}\right)\right]
    \\=& -\frac{(1-\gamma)\sin^{2}\left(\frac{\pi}{2}\frac{1-\gamma}{2}\right)}{\rho C}\Gamma\left(\frac{1+\gamma}{2}\right)\left(\frac{1}{\tau}\right)^{\frac{1+\gamma}{2}} \text{for $\tau/\beta\ll 1$}
    \\
    \chi_{loc}(\tau)=&-\frac{1}{N^2}\sum_{\sigma\sigma'}\langle T_{\tau}f^{\dagger}_{\sigma}(\tau)f_{\sigma'}(\tau)f^{\dagger}_{\sigma'}(0)f_{\sigma}(0)\rangle=G_{f}(\tau)G_{f}(-\tau)
    \\=&\frac{1}{\pi^2}C^{2}\Gamma\left(\frac{1-\gamma}{2}\right)^2\left(-\frac{1}{\tau}\right)^{1-\gamma}\\
    =&-\frac{1}{4\pi(K_{0}g_{c})^2}\frac{1}{\tau}
\end{align}
{where the last equility is valid when $\gamma$ is set to $0^{+}$.}

\section{Quantum Kinetic Equations}
Having established the equilibrium signatures of the Kondo destruction within the BFKLM, we now turn to its nonequilibrium  properties
 by analyzing 
the associated quantum transport and current fluctuations. 
We apply a finite voltage bias to the two ends of the system such that the conduction electrons experience position-dependent potentials: $V_{\text{field}}=-\sum_{i}e\bm{E}\cdot\bm{r}_{i}c^{\dagger}_{i\sigma}c_{i\sigma}$.
To realize diffusive transport, we
also add a disorder potential term to the Hamiltonian:
$H_{\text{disorder}}=\sum_{i}u_{i}c^{\dagger}_{i\sigma}c_{i\sigma}$, where $u_{i}$ is a random disorder potential whose disorder average $\overline{u_{i}}=0,\, \overline{u_{i}u_{j}}=u^2\delta_{i,j}$.

\subsection{Derivation of kinetic equations: general formulation}
The quantum kinetic equations in the disordered system are determined from Keldysh components of the Dyson equations and saddle-point equations \cite{kamenev2023field}:
\begin{align}
    [G^{-1}-\Sigma]\circ G = \hat{1}
\end{align}
\begin{equation}
    \begin{bmatrix}
   [G^{R}]^{-1}-\Sigma^{R} & -\Sigma^{K} \\
   0 & [G^{A}]^{-1}-\Sigma^{A}
   \end{bmatrix}\circ
   \begin{bmatrix}
   G^{R}(x) & G^{K}(x) \\
   0 & G^{A}(x)
   \end{bmatrix}=\hat{1}
\end{equation}
The diagonal part of the equation leads to familiar Dyson equations for retarded/advanced Green's functions. The off-diagonal part leads to quantum kinetic equations:
\begin{align}
    ([G^{R}]^{-1}-\Sigma^{R})\circ G^{K}-\Sigma^{K}\circ G^{A}=0 \label{offdiagonal}
\end{align}
where $\circ$ is the shorthand notation for:
\begin{align}
    (A\circ B)(x_{1},x_{2}) =\int d{x'}A(x_{1},x')B(x',x_{2}).
\end{align}

By defining 
\begin{align}
    G^{K}=G^{R}\circ F-F\circ G^{A}
\end{align}
and multiplying with  $([G^{R}]^{-1}-\Sigma^{R}) \, \circ$, we get
\begin{align}
    [[G^{R}]^{-1}\circ,F] = -\Sigma^{K}+(\Sigma^{R}\circ F- F\circ \Sigma^{A})
\end{align}
Performing Wigner transformations \cite{kamenev2023field}, Fourier transforming the relative coordinate part, $x-x'$, into momentum $p$, we get the quantum kinetic equations for $F(x,p)$:
\begin{align}
-\partial_{p} (\mathrm{Re}\,G^{-1})\partial_{x}F+\partial_{x}(\mathrm{Re}\,G^{-1})\partial_{p} F = -I^{\text{coll}}(\Sigma,F)
\label{kin0} 
\end{align}
where $\mathrm{Re}\,G^{-1}=G^{-1}_{0}-\mathrm{Re}\,\Sigma$ is the real part of the inverse retarded or advanced self-energy, $I^{\text{coll}}(\Sigma, F)=-i[\Sigma^{K} - F(\Sigma^{R}-\Sigma^{A})]$ is the collision integral, $x$ is the central coordinate $x=(x_{1}+x_{2})/2$ and $p$ is the momentum conjugate to the relative coordinate $x_{1}-x_{2}$. $F(x,p)$ is the Wigner transformed distribution function defined via $G^{K}(x_{1},x_{2})\equiv \int d
{x_{3}}\left[G^{R}(x_{1},x_{3})F(x_{3},x_{2})-F(x_{1},x_{3})G^{A}(x_{3},x_{2})\right]$. At equilibrium, $F(x,p)=1-2f(\omega)$ for fermions and $F(x,p)=1+2n(\omega)$ for bosons, where $f(\omega)$ and $n(\omega)$ are Fermi-Dirac and Bose-Einstein distributions, respectively.

\subsection{Quantum kinetic equations for BFKLM}
Equiped with quantum kinetic equations from Keldysh formalism, we derive the quantum kinetic equations for different degrees of freedom of BFKLM by substituting the self-energies into Eq.(\ref{kin0}).

If the propagators have no momentum dependence, $G(\omega,q)=G(\omega)$, which applies to $G_{f}$ and $G_{B}$, then at the steady state the kinetic terms vanish, leaving only:
\begin{align}
    F(\epsilon)=\frac{\Sigma^{K}(\epsilon)}{\Sigma^{R}(\epsilon)-\Sigma^{A}(\epsilon)} \, ,
\end{align}
where the distributions of local $B$ field and $f$-electrons are determined through this relation:
\begin{align}
    &n_{f}(\omega)    =\frac{\int_{\epsilon}\kappa A_{c}(\epsilon)A_{B}(\omega-\epsilon)[n_{c}(\epsilon)(1+n_{B}(\omega-\epsilon))]+g^{2}A_{\Phi}(\epsilon)A_{f}(\omega-\epsilon)[n_{f}(\omega-\epsilon)(1+n_{\Phi}(\epsilon))]}{\int_{\epsilon}\kappa A_{c}(\epsilon)A_{B}(\omega-\epsilon)[n_{B}(\omega-\epsilon)+n_{c}(\epsilon)]+g^{2}A_{\Phi}(\epsilon)A_{f}(\omega-\epsilon)[n_{\Phi}(\epsilon)+n_{f}(\omega-\epsilon)]} \, , \\
     &n_{B}(\omega)=\frac{\int_{\epsilon} A_{c}(\epsilon)A_{f}(\epsilon+\omega)n_{f}(\epsilon+\omega)[1-n_{c}(\epsilon,x)]}{\int_{\epsilon} A_{c}(\epsilon)A_{f}(\epsilon+\omega)[n_{f}(\epsilon+\omega)-n_{c}(\epsilon)]} \, ,
\end{align}
where $n_{f}(\omega)=[1-F_{f}(\omega)]/2$, 
and $n_{B}(\omega)=[F_{B}(\omega)-1]/2$. Here 
the dependence on the position 
variable $x$ is hidden for convenience. We have incorporated $\kappa=1/N$ corrections to the $f$-electron distribution.
The kinetic equations for the conduction electrons and vector bosons $\bm{\Phi}$ are determined by
\begin{align}
     &\partial_{k}\epsilon_{k}\partial_{x}n_{c}(\bm{k},\epsilon,x)+eE\partial_{k}n_{c}(\bm{k},\epsilon,x)=-I_{imp}^{\mathrm{coll}}(\bm{k},\epsilon,x)-I_{B}^{\mathrm{coll}}(\bm{k},\epsilon,x) ,\\
     &-\partial_{q} G^{0}_{\Phi}(\bm{q},\omega)^{-1}\partial_{x}n_{\Phi}(\omega,x)=-I_{\Phi f}^{\mathrm{coll}}(\omega,x)   \, ,
\end{align}
with the electron collision integrals $I_{imp}$ and $I_{B}$, arising from scattering with impurities and scattering with localized $f$-electrons mediated by $B$ field at the Kondo-destruction QCP, and a bosonic collision integral describing scattering between bosons and localized $f$-electrons:
\begin{align}
    I_{imp}^{\mathrm{coll}}
    =&u^{2}\sum_{k'}A_{c}(\bm{k}',\epsilon)[n_{c}(\bm{k},x,\epsilon)-n_{c}(\bm{k}',x,\epsilon)]
    =\frac{n_{c}(\bm{k},\epsilon,x)-n_{c}(\epsilon,x)}{\tau},\\
    I_{B}^{\mathrm{coll}}=&\frac{1}{N}\int{\frac{d\omega}{2\pi}}A_{B}(\omega,x)A_{f}(\omega+\epsilon)\left\{n_{c}(\bm{k},\epsilon,x)[1-n_{f}(\epsilon+\omega)]n_{B}(\omega,x) -[1-n_{c}(\bm{k},\epsilon,x)]n_{f}(\epsilon+\omega)[1+n_{B}(\omega,x)]
   \right\},\\
   I_{\Phi f}^{\mathrm{coll}}=&\frac{g^{2}}{N}\int \frac{d\epsilon}{2\pi}\,A_{f}(\epsilon+\omega)A_{f}(\epsilon)\left\{n_{\Phi}(\omega)[1-n_{f}(\epsilon)]n_{f}(\omega+\epsilon)-[1+n_{\Phi}(\omega)]n_{f}(\epsilon)[1-n_{f}(\omega+\epsilon)]\right\},
\end{align}
where $n_{c}$ and $n_{\Phi}$ are the distribution functions of the conduction electrons and vector bosons, respectively. $\frac{1}{\tau}=2\pi\rho u^2$ is the impurity scattering rate, $n_{c}(\epsilon,x)$ is the even sector of the nonequilibrium distribution function $n_{c}(\bm{k},\epsilon,x)=n_{c}(\epsilon,x)+\delta n_{c}(\bm{k},\epsilon,x)$, which is momentum-independent due to impurity scattering. $\delta n_{c}(k,\epsilon,x)$ captures the odd deviation, and is crucial in determining electric current.

We find that the vector bosons $\Phi$ and $f$-electrons form a closed system and are only away from equilibrium at the order of $O(1/N^{2})$ and $O(1/N)$, respectively; therefore, when the Kondo effect is destroyed, at leading order in $1/N$, they remain in equilibrium in the presence of the applied electric voltage: $n_{\Phi}(\omega)=\frac{1}{e^{\beta\omega}-1},n_{f}(\epsilon)=\frac{1}{e^{\beta\epsilon}+1}$ at zero temperature. 
However, the hybridization field $B$, being a charged composite of $f$ and $c$ electrons, is driven out of equilibrium along with the current-carrying conduction $c$ electrons at leading order.

The spatial profile of the even part is governed by the diffusion equation:
\begin{align}
    D \partial_x^2 n_{c}(\epsilon, x) = I_{ee}^{\mathrm{coll}}(\epsilon, x) + I_{B}^{\mathrm{coll}}(\epsilon, x)
\end{align}
Here $D=\tau v_{F}^{2}/d$ is the diffusion constant,  $I_{ee}^{\mathrm{coll}}(\epsilon,x)$ is the electron-electron collision integral introduced for completeness.
In the regime where inelastic scattering is strong, $n_{c}(x,\epsilon)=\frac{1}{e^{\frac{\epsilon-\mu(x)}{T_{e}(x)}}+1}$, where $\mu(x)=-eEx$. $T_{e}(x)$ denotes the local electronic temperature induced by applied voltage bias, subject to the boundary conditions: $T_{e}(\pm L/2)=0$.
The nonequilibrium hybridization $B$ field has the following 
distribution function and Green's function:
\begin{align}
    n_{B}(\omega, x) =& 
\frac{-\mathrm{Li}_{\frac{1}{2}}\left[-e^{-\frac{\omega + \mu(x)}{T_{e}(x)}}\right]}
     {\mathrm{Li}_{\frac{1}{2}}\left[-e^{-\frac{\omega + \mu(x)}{T_{e}(x)}}\right] 
     - \mathrm{Li}_{\frac{1}{2}}\left[-e^{\frac{\omega + \mu(x)}{T_{e}(x)}}\right]}\\
     G_{B}^{R}(\omega,x)=&\frac{\sqrt{2}T_{e}(x)^{\frac{1}{2}}/(\rho\sqrt{\pi})}{e^{i\frac{\pi}{4}}\mathrm{Li}_{\frac{1}{2}}\left[-e^{-\frac{\omega+\mu(x)}{T_{e}(x)}}\right]+e^{-i\frac{\pi}{4}}\mathrm{Li}_{\frac{1}{2}}\left[-e^{\frac{\omega+\mu(x)}{T_{e}(x)}}\right]}
\end{align}
where $\mathrm{Li}_{\alpha}(x)$ is the polylogarithm function and $\Gamma(x)$ is the gamma function.
Multiplying $\epsilon$ and $L^2$ to both sides and integrating out $\epsilon$ 
leads to
\begin{align}
    L^{2}\frac{d^{2}T_{e}^{2}}{dx^2}+\frac{6}{\pi^2}(eV)^{2}=\frac{L^2}{l_{B}^{2}}T_{e}^{2}
\end{align}
where $l_{B}$ is the scattering length between the conduction electrons and 
$f$-electrons mediated via the nonequilibrium critical $B$-field.
The solution of $T_{e}$ is given by
\begin{align}
    T_{e}(x)=\frac{\sqrt{6}\,(eV)l_{B}}{\pi L}\sqrt{1-\frac{\cosh{(x/l_{B})}}{\cosh{(L/2l_{B})}}}
\end{align}
shown in the main text.

\end{document}